\documentclass[aps,prl,twocolumn,groupedaddress]{revtex4-1} 
\usepackage{graphicx}
\usepackage{subfigure}
\usepackage{bm}

\begin{document}

\title{Degenerate Fermi Gas of $^{87}$Sr}
\author{B. J. DeSalvo, M. Yan,  P. G. Mickelson, Y. N. Martinez de Escobar,   and T. C. Killian}
\affiliation{Rice University, Department of Physics and Astronomy, Houston, Texas, 77251}

\date{\today}

\begin{abstract}
We report quantum degeneracy in a gas of ultra-cold fermionic $^{87}$Sr atoms.  By evaporatively cooling a mixture of spin states in an optical dipole trap for 10.5\,s, we obtain samples well into the degenerate regime with $T/T_F=0.26^{+.05}_{-.06}$.  The main signature of degeneracy is a change in the momentum distribution as measured by time-of-flight imaging, and we also observe a decrease in evaporation efficiency below $T/T_F \sim 0.5$.

\end{abstract}

\pacs{03.75.Ss}

\maketitle

Quantum degeneracy of fermions in dilute atomic gases \cite{gps08} is currently one of the most active areas of physics research.
Tunable interactions \cite{cgj09} have allowed an exploration of the BEC-BCS crossover \cite{bdz08} as well as  study of phenomena in the unitary regime such as superfluidity in systems with \cite{plk06,zss06} and without \cite{ohg02,cba04Science,zas05} spin imbalance.
Research on Fermi gases in optical lattices \cite{kms05} can make direct connections to the properties of electrons in solids and realize important models like the Hubbard Hamiltonian \cite{jsg08}, but with new capabilities for controlling system parameters such as density, interaction strength, and dimensionality.
 Current searches are underway for analogues of high temperature superconductivity \cite{hcz02} and spontaneous N\'eel magnetism \cite{ddl03}.

Here we report quantum degeneracy of the fermionic isotope of strontium, $^{87}$Sr. Five other species of fermions have been brought into the quantum degenerate regime ($^{40}$K \cite{dji99}, $^{6}$Li \cite{tsm01}, $^{3}$He \cite{mjt06}, $^{171}$Yb and $^{173}$Yb \cite{ftk07}), but $^{87}$Sr has several properties which make quantum degenerate samples of this type particularly interesting. $^{87}$Sr has a very large nuclear spin, I=9/2, which may allow studies of novel magnetic phenomena due to enlarged SU(N) symmetry of the interaction Hamiltonian for $N=2I+1$ \cite{ghg10,hgr09,chu09}. High resolution spectroscopy technologies are the most advanced in strontium because of the use of narrow intercombination transitions in $^{87}$Sr for optical frequency standards \cite{ykk08}, and these tools have motivated proposals for applications in quantum information \cite{dby08,grd09,rjd09} and quantum simulation of many-body phenomena \cite{ghg10,fhr09}.  Strontium also has a number of stable bosons which have recently been brought into the quantum degenerate regime \cite{sth09,mmy09,mmy10}, which makes Fermi-Bose mixtures with relatively small mass differences available \cite{ide06}.
There is also the potential for manipulating interactions on small spatial and temporal scales with low-loss optical Feshbach resonances \cite{ctj05,ekk08}.

Details about our apparatus can be found in \cite{mmp08,mmy09,mma09}.   Atoms are trapped from a Zeeman slowed beam in a magneto-optical trap (MOT) operating on the $(5s^2)^1S_0-(5s5p)^1P_1$ transition at 461\,nm.  Since this transition is not closed, approximately 1 in $10^5$ excitations results in an atom decaying through the $(5s5d)^1D_2$ state to the $(5s5p)^3P_2$ state,
 which has an 9 minute lifetime \cite{yka04} and can be trapped in the quadrupole magnetic field of the MOT \cite{fdp06,nsl03,xlh03,pdf05}.  The magnetic trap has a lifetime of about 25\,s, which is limited by background pressure and blackbody radiation \cite{xlh03}. This allows us to accumulate  atoms over a period of 30\,s and trap a significant number in spite of the low natural abundance of $^{87}$Sr (7\%).

After accumulation, atoms in the $^3P_2$ state are returned to the ground state via excitation to the $(5s4d)^3D_2$ state \cite{mma09}.  This repumping is achieved by applying 3\,W/cm$^2$ of 3.012 \,$\mu$m light for 60\,ms.  The transition has hyperfine structure due to the nuclear spin of $I=9/2$, which spreads the transition over $\sim$3\,GHz. Individual hyperfine transitions are not resolved in the repumping efficiency curve because of the high intensity of the repump laser and length of time over which the repumping laser is applied \cite{mic10}.
The 3\,$\mu$m laser is tuned 1\,GHz blue of the $^{87}$Sr centroid, which effectively repumps transitions from the  $F=11/2$ and  $F=13/2$ $^3P_2$  levels.  During the repumping stage, the 461\,nm MOT is left on so that atoms returned to the ground state are recaptured and cooled.  We typically recapture $3 \times 10^7$ atoms at temperatures of a few milliKelvin. It should be possible to improve this number  by modulating the repumping laser frequency  to excite atoms in all of the $F$ levels.

\begin{figure}

\includegraphics[keepaspectratio=true,width=3.3in]{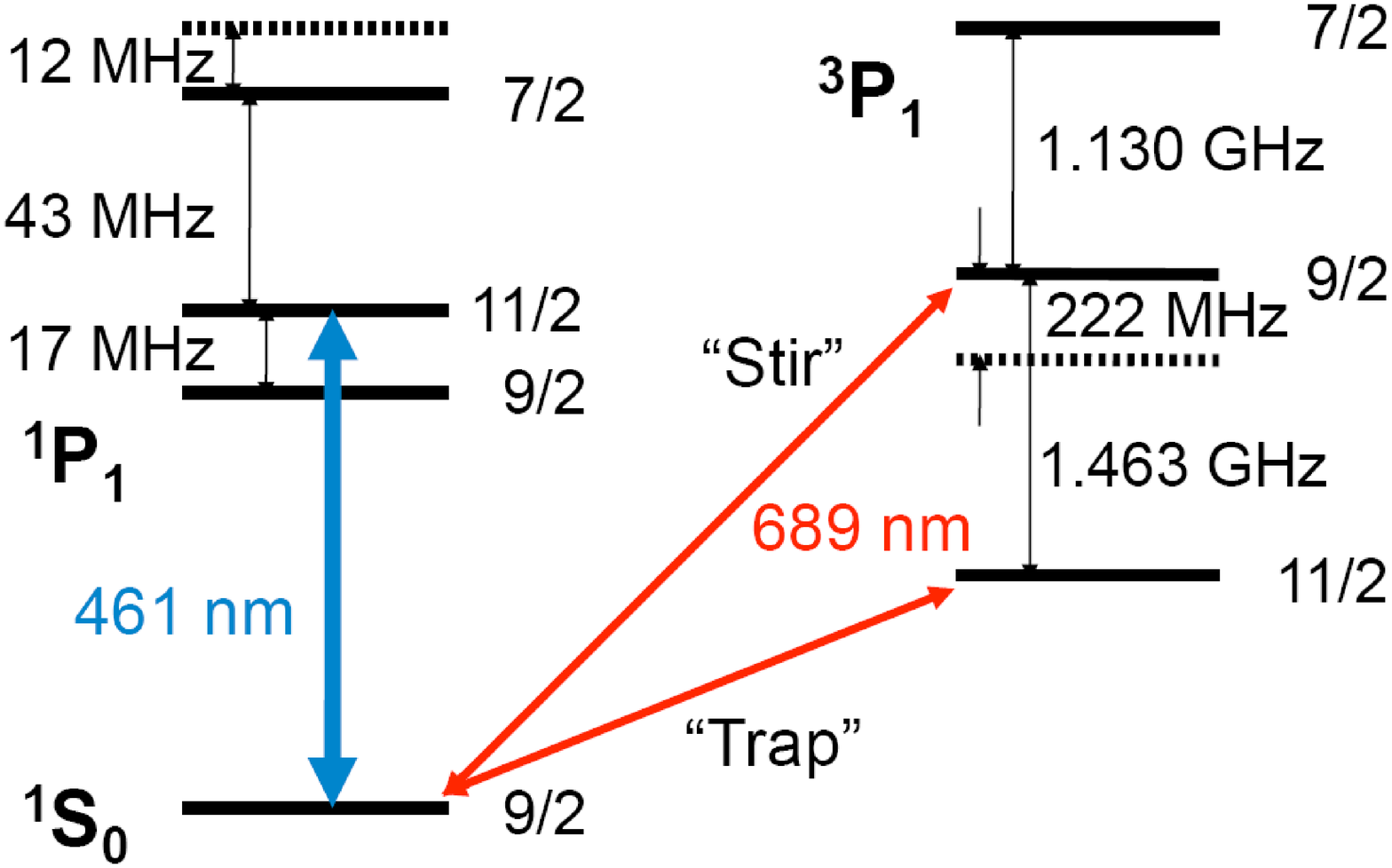}\\
\caption{(color online) Partial level diagram for $^{88}$Sr (- -) and $^{87}$Sr (---) including hyperfine structure and isotope shifts.  For $^{87}$Sr, total quantum number $F$ is indicated. }
\label{Levels}

\end{figure}

Following this step, the 461\,nm light is extinguished and  689\,nm light is applied to drive the $(5s^2)^1S_0-(5s5p)^3P_1$ transition and create an intercombination line MOT.  Similar to \cite{mki03}, two frequencies are applied (Fig. \ref{Levels}).
The two frequencies are required because the
Zeeman shift is much larger in the excited state than the ground state, causing some ground-state $m_F$ levels to be anti-trapped.
A ``trap laser", slightly red-detuned from the $F=9/2 \rightarrow F=11/2$ hyperfine transition, creates a stable trapping force on average for atoms only in the presence of a ``stir laser" red-detuned from  the $F=9/2 \rightarrow F=9/2$ transition. The stir laser effectively randomizes the populations
because the Zeeman shift for this transition is much smaller than for the trapping transition. For a strong transition, such as the $^1S_0-^1P_1$ transition at 461\,nm, this mixing is accomplished by a single trapping laser, but an extra laser is required for the intercombination transition because excitation is relatively slow.

To increase the initial capture rate from the 461\,nm MOT,  both the stir laser and trap laser are artificially broadened so that they are resonant with a wider velocity range of atoms.  This is achieved by modulating the frequencies of the acousto-optic modulators generating the beams at a frequency of 30\,kHz with a peak-to-peak amplitude of 660\,kHz. Initial detunings are -1.2\,MHz and -900\,kHz from the stir and trap transitions respectively, and intensities are 1\,mW/cm$^2$ and 2.2\,mW/cm$^2$.   This allows trapping of up to 70\% of the atoms in the intercombination-line MOT.  During a 400\,ms cooling time, the detuning, dither, and intensity are reduced to -90\,kHz, 200\.kHz, and 0.8\,mW/cm$^2$ for the trap laser and -600\,kHz, 200\,kHz and 0.9\,mW/cm$^2$ for the stir laser.  Simultaneously, the magnetic field gradient is increased from 0.2\,G/cm to 1.9\,G/cm. This increases spatial confinement and transfer to the optical dipole trap (ODT).

After the cooling and compression stage, an ODT, consisting of two crossed laser beams, is overlapped for 100\,ms with the intercombination-line MOT with a modest power (3.9 W) per beam. A single beam derived from a 21\,W multimode, $1.06$\,$\mu$m fiber laser is recycled through the chamber to form the ODT.  The resulting trap has equipotentials that are nearly oblate spheroids, with the tight axis close to vertical.  Within the trapping region, each beam has a waist of approximately $90$\,$\mu$m.
During this loading time, the dither amplitudes are reduced to zero, and the intensities and detunings of the MOT beams are further reduced to approximately 3\,$\mu$W/cm$^2$ and  -30\,kHz.

After the extinction of the 689\,nm light, the ODT power is ramped in 30\,ms to 7.5\,W per beam to obtain a trap depth of $25$\,$\mu$K.  At this point, we typically trap $3 \times 10^6$ atoms at 7\,$\mu$K with a density of $2.5 \times 10^{13}$\,cm$^{-3}$.  No attempt is made to spin polarize the sample. Experiments using $^{87}$Sr for optical frequency standards \cite{lud08} found that the intercombination-line MOT produced a sample with roughly equal populations of ground-state magnetic sublevels.  For a conservative estimate of the phase-space density, we will assume equal populations in all levels. This corresponds to an average initial collision rate of 200\,s$^{-1}$ and a phase-space density of approximately $10^{-3}$. As discussed below, exact determination of the polarization is not necessary to establish Fermi degeneracy of our sample and will be the topic of future studies. The sample lifetime in a static ODT, presumably limited by background gas collisions, is 30\,s, which is sufficient for evaporative cooling.

After loading into the ODT, we begin forced evaporation by lowering the power in the ODT according to the formula $P= P_0/(1+t/\tau)^\beta  + P_{\mathrm{offset}}$ with time denoted by $t$, $\beta=1.4$, and $\tau=1.5$\,s.  This trajectory was designed \cite{ogg01} without $P_{\mathrm{offset}}$ to yield efficient evaporation when the effect of gravity can be neglected.  Gravity is a significant effect in this trap for Sr, and to avoid decreasing the trap depth too quickly at the end of the evaporation, we set $P_{\mathrm{offset}} = 0.7$\,W which corresponds to the power at which gravity causes the trap depth to be close to zero.

For diagnostics, the ODT laser is extinguished, and atoms are allowed to expand for a time-of-flight (TOF) of between 20\, and 22\,ms. We then obtain absorption images using a linearly polarized laser on resonance with the $^1S_0(F=9/2)-{^1P_1}(F=11/2)$ transition. The intensity is $0.02I_s$, where $I_s=42$\,mW/cm$^2$ is the saturation intensity, and the exposure time is 35\,$\mu$s. Due to the small splitting between the $F=11/2$ and $F=9/2$ $^1P_1$ states, the imaging beam excites transitions to both states. For a given density of atoms, the optical depth of the sample depends upon the population distribution of ground state magnetic sublevels. Numerical solution of the optical pumping rate equations show that atoms quickly approach a steady state distribution. For the small initial  polarization in our samples the variation in optical depth is relatively small and an average absorption cross section of $\sigma=3 \lambda^2/4\pi$, where $\lambda=461$\,nm, is accurate to within 10\%. This diagnostic does not provide enough information to distinguish atoms in different magnetic sublevels.

\begin{figure}
\includegraphics[keepaspectratio=true,width=3.5in,height=2.7in]{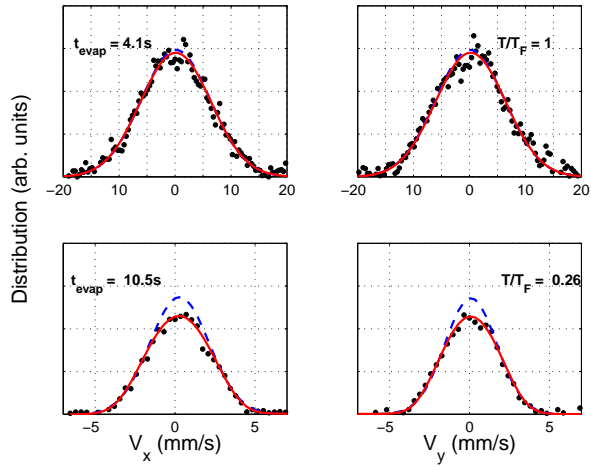}\\
\caption{(color online) Velocity distributions along axes perpendicular to imaging beam. At $t=4.1$\,s of forced evaporation (top), the data is fit well with a Maxwell-Boltzmann(- -) or  Fermi-Dirac distribution(--). At 10.5\,s, classical-particle statistics overestimates the population at low velocities, while Fermi-Dirac statistics for $T/T_F=0.26$ accurately fit the data.}
\label{Raw Data}
\end{figure}

Figure \ref{Raw Data} shows data and fits for one-dimensional slices through the center of TOF absorption images along  axes perpendicular to the imaging beam and to each other. Position has been converted to velocity assuming ballistic expansion and long expansion time. Ballistic expansion is a good approximation because the mean field interaction energy is negligible compared to the kinetic energy, and the expansion times used are much longer than the relevant time scale, $1/\omega$, where $f=\omega/2\pi \approx 100$\,Hz is the typical mean trap oscillation frequency.

Well outside the quantum degenerate regime, the Maxwell-Boltzmann and Fermi-Dirac momentum distributions are not distinguishable.
To fit the full TOF images accurately when the atoms are in the quantum degenerate regime, we must use the Fermi-Dirac expression for the two-dimensional momentum distribution \cite{dem01}
\begin{equation}
\pi(p_x,p_y)=-\frac{(k_B T)^2}{2\pi m (\hbar \omega)^3} Li_2[-\zeta e^{-\frac{p_x^2+p_y^2}{2mk_BT}}]
\label{FD fit}
\end{equation}
where $Li_n$ is the  poly-logarithmic function of order $n$, $p_{x/y}$ are momenta transverse to the imaging beam, and $\zeta$ is the fugacity.
The fugacity is related to the sample temperature ($T$) and Fermi temperature ($T_F$)  by $Li_3[-\zeta]=-\frac{1}{6(T/T_F)^3}$, so this method yields a measure of the quantity $T/T_F$ without an explicit assumption on the spin polarization. Formally, the signal is a sum of distributions for each magnetic sublevel, where each may have different fugacities. Because of the weak dependence of $T_F=(6 N)^{1/3} \hbar \omega / k_B$ on the number of atoms in a single quantum state ($N$),  atoms in magnetic sublevels with the smallest $T/T_F$ values will dominate the signal.  Sublevels with significantly lower population and higher values of $T/T_F$ will contribute less to the signal, but they do cause the extracted value of $T/T_F$ to be an upper bound of the value for the most populated state.
The weak dependence of the velocity distribution on the fugacity far from degeneracy makes extracting $T/T_F$ difficult for hotter clouds.

When fitting the TOF images using Maxwell-Boltzmann statistics, the entire image is  fit with a Gaussian.  This fit determines the number of atoms \cite{dem01} and the size of the cloud, $s$, where the density is $n \propto \mathrm{exp}(-r^2/2s^2)$.  To determine the temperature, the high velocity wings of the distribution, consisting only of atoms at radii greater than  $s$, are fit to a Maxwell-Boltzmann velocity distribution.  The deviation of the full distribution from a Gaussian is never large and the high velocity wings are less sensitive to the changes in the shape of the distribution as the atoms become degenerate \cite{dji99}. Numerical simulations of velocity distributions show that this method of thermometry is accurate to within 10\% for $T/T_F>0.4$ and becomes less accurate for more degenerate samples. The temperature is always overestimated.  Combining this measure of temperature with knowledge of the ODT oscillation frequencies provides a measurement of $T/T_F$. Here we assume equal population of the magnetic substates, which makes this an upper bound on $T/T_F$ for the magnetic sublevel with the largest population.

After 4.1\,s of forced evaporation, the sample has reached $T/T_F=1$ as measured by both methods of fitting, although the uncertainty is large for the Fermi-Dirac fitting.  Both the Maxwell-Boltzmann and the Fermi-Dirac fitting describe the data well (Fig.\ \ref{Raw Data}).  After 10.5\,s the Maxwell-Boltzman distribution overestimates the number of atoms with small velocity in the central region of the cloud.  However, the Fermi-Dirac fitting yields an excellent description of the data with a fugacity corresponding to $T/T_F=0.26^{+.05}_{-.06}$.  The non-Gaussian character of the distribution is a clear signature of the onset of quantum degeneracy and the limiting of occupancy of lower energy levels due to the Pauli exclusion principle.

\begin{figure}
\includegraphics[keepaspectratio=true,width=3.5in,height=2.7in]{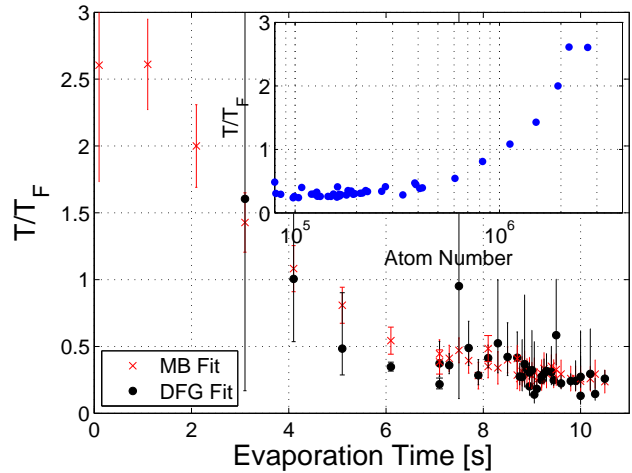}\\
\caption{(color online) Evaporation of $^{87}$Sr. $T/T_F$ is obtained from either a fit of the velocity distribution to a single Fermi-Dirac distribution (DFG), or a determination of number and temperature from fits to a Maxwell-Boltzmann distribution and knowledge of the trap oscillation frequencies (MB). (Inset) Variation of $T/T_F$ with atom number. $T/T_F$ is calculated using the latter of the two methods.  Evaporation is efficient early in the trajectory and $T/T_F$ decreases significantly as atoms are lost. But $T/T_F$ plateaus near the limit of about $\sim 0.25$ as the evaporation efficiency decreases.}
\label{TTF vs Time}
\end{figure}

Figure \ref{TTF vs Time} shows the evolution of  $T/T_F$ during the forced evaporation, determined by the two methods described above. The temperature starts at 4\,$\mu$K and reaches 30\,nK after 10\,s of evaporation.  We achieve a limiting value $T/T_F =0.25$ with this evaporation trajectory, and the agreement of the two methods of determining this quantity indicate that the assumption of equal ground state sublevel populations is a reasonable approximation for this data.

Figure \ref{TTF vs Time} (inset) provides information on why lower values of $T/T_F$ are not observed. Evaporation initially proceeds efficiently through collisions of $^{87}$Sr atoms in different magnetic sublevels. $T/T_F$ drops by a factor of almost 10 for a loss of about a factor of 10 in atom number.
  At values below $T/T_F =0.5$, the evaporation becomes very inefficient.  This signature has been seen in evaporation of $^{40}$K \cite{dji99} and $^{173}$Yb \cite{ftk07} and ascribed to Pauli blocking of collisions when atoms have reached the quantum degenerate regime.

We have described the creation of a quantum degenerate Fermi gas of $^{87}$Sr.  This result opens possibilities of future studies involving quantum degenerate Bose-Fermi mixtures \cite{mmy09,mmy10},  quantum degenerate Fermi gases with large ground state degeneracy \cite{ghg10,hgr09,chu09}, and quantum computing with alkaline-earth metal atoms \cite{dby08,grd09,rjd09}.

\textmd{\textbf{Acknowledgements}}
 This research was supported by the Welch Foundation
(C-1579), National Science Foundation (PHY-0855642), and the Keck Foundation.

\end{document}